# WAVEFRONT SHAPING OF THE PUMP IN MULTIMODE FIBER AMPLIFIERS: THE GAIN-DEPENDENT TRANSMISSION MATRIX

Tom Sperber[1,2], Vincent Billault[2], Patrick Sebbah[1,3], Sylvain Gigan[2]

[1] *ESPCI ParisTech, PSL Research University, CNRS, Institut Langevin, 1 rue Jussieu, F-75005, Paris, France*

[2] *Laboratoire Kastler Brossel, ENS-PSL Research University, CNRS, UPMC-Sorbonne universités, Collège de France ; 24 rue Lhomond, F-75005 Paris, France*

[3] *Department of Physics, Jack and Pearl Resnick Institute for Advanced Technology, Bar-Ilan University, Ramat-Gan, 5290002 Israel*

**Abstract**

Wavefront shaping techniques allow the control of the transport of light through many types of scattering or complex media, among them multimode fibers. The case of a multimode fiber which is also a gain medium presents further intriguing prospects for control, due to the possibility of wavefront shaping not only the signal light amplified by the medium, but also the pump light absorbed within it. Provided that the lightwave used for pumping is coherent, its shaping prior to injection will affect the complex, speckle-like spatial patterns of excitation within the amplifier volume, which in turn will act upon the signal as heterogeneous gain. We introduce a new theoretical model which captures the essential features of a multimode amplifying fiber pumped by a coherent beam with configurable modal content, allowing the calculation of the gain-dependent transmission matrix. We numerically implement our model to explore the extent to which one may control the amplified light in the spatial-domain, by shaping the pump; we demonstrate a significant ability of manipulation, as well as several interesting physical mechanisms limiting it. In particular, we show how taking into account the cross-contributions between all guided modes within the fiber is essential for understanding the full range of the signal sensitivity to the choice of pump shaping.

**Introduction**

The spatial modulation of a coherent beam prior to its coupling into a multimode fiber (MMF) enables precise tailoring of the light propagation within it. The intense research interest shown in recent years has dealt mostly with passive fibers [1-3], where attention is focused on the control of the complex intensity patterns emerging at the output. If a gain medium is also present in the system, in particular considering the pump beam in a multimode fiber amplifier (MMFA), the intensity patterns all along the propagation path within the fiber core are of interest, since they may form spatially complex gain profiles acting upon the signal. The prospect of spatial modulation of the pump beam prior to its injection into the fiber should open intriguing avenues of indirectly tailoring the amplified light. Fig. 1 schematically depicts such a system, essentially a MMFA with a pump beam subject to wavefront shaping (WFS) by means of a spatial light modulator (SLM).

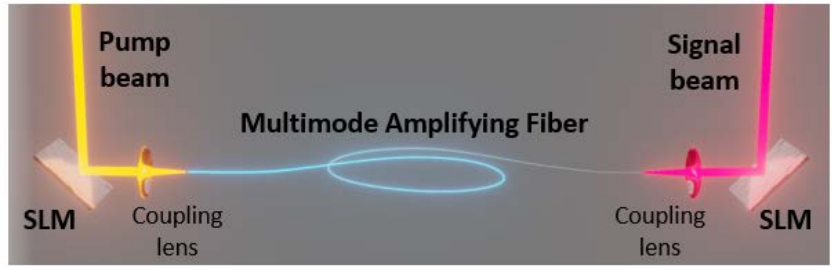

Figure 1 - Schematic illustration of the studied system.

The motivations to study the transmission of a MMFA and specifically its dependence upon a configurable pump scheme are numerous. Interest in fiber amplifiers operating in the multimodal regime has considerably risen in recent years in several applicative domains: In the context of long-haul optical telecommunication links, where amplifiers are crucial in-line repeaters [4], the imminent exhaustion of the information capacity of singlemode links drives the development of Spatial Domain Multiplexing (SDM) schemes [5-6]. In the fields of laser sources for industrial applications such as marking or machining, as well as for sensing applications such as Lidar, the ever-present motivation to increase the output power has traditionally pushed towards fibers with large core diameters in order to avoid non-linear effects [7-8]; however the resulting growth in the number of guided modes is largely undesired, and various design approaches and measures are taken to suppress the high-order modes so as to maintain good beam quality [9-11]. Furthermore, from the perspective of less applicative and more fundamental research within the sphere of the control of light in disordered media, this type of system holds a unique quality: whereas the 'traditional' shaping of the signal enables control of its propagation through some given, fixed disorder (in this case, dictated by the multimodal waveguide), the shaping of the *pump* offers the intriguing opportunity to tailor the disorder itself.

Recent works seeking to exploit the fact that the pump in a MMFA is also multimodal and may therefore be spatially shaped, have been carried out largely within the domain of optical telecommunications. The interest in this field has been limited to the requirement of "gain equalization", i.e. the minimization of the 'modal dependent gain', by compensating for the natural tendency of the higher-order spatial modes to suffer greater losses in the long-haul link, as well as lower gain values within the amplifier [12]. Successful equalization by manipulation of the pump's modal content, e.g. using phase plates [13-14] or an SLM [15], has been demonstrated. The number of guided signal modes was typically limited to 4-6 modes, since this is the range relevant for the framework of SDM applications. For the same reason, only the absolute amplification values of the different modes were considered; the aspects relevant to the spatial-domain observation of the amplifier output, including the gain-dependent phases, were naturally not examined. In contrast, recent studies of MMFAs within the context of sources for high-power industrial/sensing applications, have certainly explored spatial-domain aspects; however, the wavefront shaping techniques within them have been applied only upon the signal beam, while using incoherent and non-modulated pump schemes [16-18].

In order to fully describe the transport of the signal light through the fiber amplifier, we employ the notion of the transmission matrix (TM), as defined in [19] – the linear operator relating the optical field at the output to that at the input. Recent research works have applied this formalism to study light propagation through passive (i.e. non-amplifying) fibers, demonstrating remarkable advances in the ability to connect experimental TM measurements to theoretical models based on the finite set of waveguide modes [20-21]. For our amplifying system, we confine ourselves to the undepleted pump regime so that the transmission remains linear in the signal field, and the TM formalism may be applied.

In this work we present, for the first time to the best of our knowledge, a complete model for the TM of a coherently pumped MMFA, as a function of the modal content of the injected pump beam. We report on the results of a numerical simulation of this model, quantifying the degree to which pump shaping may be used to control the amplifier output in the spatial domain.

**Results**

*Proposed Model for the amplifier TM*

Starting with a description of the propagation of light in a multimode fiber in the absence of gain or absorption, we first choose a set of transverse modes $\psi_i$, $i = 1 \ldots N$ which best represent the fiber's eigenmodes. This essentially means that the modes propagate with negligible coupling to each other, changing only the phase terms of their complex amplitudes. Therefore, the optical field at any point along the fiber's axis of propagation z is written as:

$$E(z) = \sum_i^N Ain_i * e^{-j\beta_i z} * \psi_i (r,\theta) \qquad (1)$$

where $(r,\theta)$ are the transverse coordinates, $\beta_i$ is the propagation constants of eigenmode $\psi_i$, and $Ain_i$ is the complex amplitude with which each eigenmode was launched into the fiber at the injection point at z=0. Stated in matricial terms, eq. (1) simply means that the TM of the fiber (expressed in the modal basis) is readily given as a diagonal matrix with the phases along the diagonal determined by the propagation velocities and the fiber length. The basis used in our numerical simulations was the set of Linearly-Polarized (LP) modes well-known in the textbook literature of MMFs [22-23], however our theoretical model is hereby presented without loss of generality, and any set of orthogonal modes chosen to approximate the waveguide's base of eigenmodes may be plugged in. As was recently demonstrated in [21], for propagation distances on the order of centimeters, the LP modes may serve with excellent precision as such an approximation, and the set of circularly polarized (CP) modes may be used for more precise modeling of the propagation over a longer fiber. Fig. 2 serves as a visualization of the intensity evolution that arises along the propagation due to the modal dephasing.

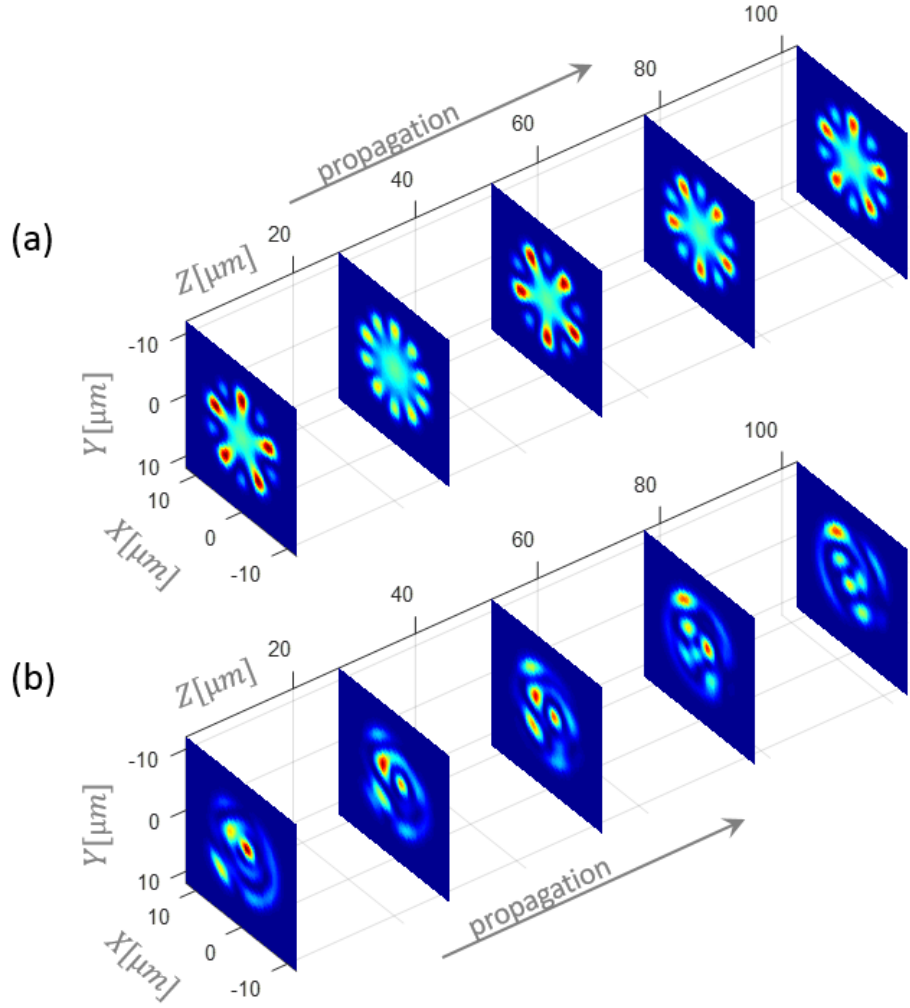

Figure 2 - examples of 2 possible pump configurations propagating through the fiber: (a) Cross-sections showing the pump's intensity patterns in the case of only 2 modes (LP01 and LP51) excited. The propagation causes a periodic beating of these modes; (b) same cross-sections for a case where all the ~20 supported pump modes are excited. The propagation causes a more complicated evolution of the intensity.

Eq. (1) provides a fully deterministic prediction of the evolution of the optical field along the fiber, once the set of values of all complex amplitudes $Ain_i$ – which we refer to as the *modal content,* or *configuration* – has been specified. In order to numerically simulate this evolution, we model the fiber as a series of truncated thin 'slices', i.e. segments of a fixed length (which we shall denote as $\Delta z$) chosen to be much smaller than the typical dephasing length $2\pi/(max\{\beta_i\} - min\{\beta_i\})$. This means $\Delta z$ is sufficiently short (typically in the fiber studied, on the order of a few microns) so that the field in each slice may be considered to be constant within it.

Since the eigenmode set is a base for any guided lightfields, the injection amplitudes are determined as the coefficients of decomposition upon this base, of the optical field injected at the fiber facet:

$$Ain_i = \iint E_{incident}(z=0) \cdot \psi_i^* dr d\theta \tag{2}$$

Gain is now added to the medium in the form of a rare-earth element doping, spread throughout the fiber core with uniform doping concentration $N_d$. We follow the standard treatment in the literature (for example [24-25]) where the electronic levels are modeled as a two-level gain system, with $n_1, n_2$ representing, respectively, the fractional lower and upper level populations. The 'pump' wavelength $\lambda_{pump}$ is the lightwave efficiently exciting the upper level population, and is guided by the modes $\psi_i^p$; while stimulated emission from this excitation efficiently amplifies the 'signal' wavelength $\lambda_{sig}$, with the corresponding set of modes $\psi_i^s$. Lastly, we work under the assumption of weak signal power with respect to the pump power (commonly known as non-depletion of the pump). In this regime, the rate equations are easily solved (see [25]) by considering only the pump intensity $I_p$ arriving at any given point:

$$n_2 = \frac{\sigma_a^{pump} \cdot I_p}{(\sigma_a^{pump} + \sigma_e^{pump}) \cdot I_p + hc/(\tau \lambda_{pump})} \tag{3}$$

where $\sigma_a^{pump}, \sigma_e^{pump}$ are, respectively, the absorption and emission cross-sections at the pump wavelength, and $\tau$ is the lifetime of the upper level excitation.

Any optical intensity $I$ propagating through the medium is amplified/absorbed by the upper/lower levels, respectively, with efficiencies determined by the (wavelength-dependent) emission/absorption cross-sections $\sigma_e, \sigma_a$ thus:

$$\frac{dI}{dz} = (\sigma_e n_2 - \sigma_a n_1) N_d \cdot I \tag{4}$$

Expressed in terms of field $E$ rather than intensity, the interaction with the levels may be written as a modification of the medium's refractive index, rendering it complex: $n = n_0 + n' + j \cdot n''$, where $n_0$ is the index in the absence of the gain element, $n''$ contributes the amplitude change (gain or absorption), and therefore satisfies $n'' = \frac{1}{2}(\sigma_e n_2 - \sigma_a n_1) N_d$, and $n'$ determines the associated (gain-induced) change of the wave's phase. This relation holds for both pump and signal wavelengths, if the corresponding cross-sections are plugged in. The real and the imaginary parts of the complex susceptibility are linearly related through the Kramers-Kronig relation [26], and the factor relating the two parts, often referred to as the Refractive Index Change (RIC) parameter $K_{RIC} = n'/n''$, depends on the medium and its specific electronic level structure. Values may be found in the literature for specific media [27-29]. Taking into consideration eq. (3) and (4), we may express the optical field that has propagated a very short distance $\Delta z$ through the medium, as:

$$E(z + \Delta z) = E(z) \cdot e^{-j\beta_0 \Delta z} \cdot \left\{1 - j\beta_0 \Delta z \cdot \tfrac{1}{2}(\sigma_e n_2 - \sigma_a n_1)N_d \cdot (j + K_{RIC})\right\} \quad (5)$$

because the gain-induced changes to the refractive index are small enough to be approximated using the first-order Taylor expansion of the exponential function: $e^{x+\varepsilon} \approx e^x(1+\varepsilon)$.

We now confront the central question of our system: how does the signal field interacting with a population excitation with some arbitrary, 'speckly' spatial distribution (because it has been generated by the multimodal pump's intensity) propagate through the fiber? It is evident from eq. (5) that once the signal field has been amplified along some given fiber slice, it is no longer necessarily a perfectly guided field; the stimulated emission contribution which it has picked up mirrors the complicated pump spatial distribution. The approach at the heart of our model is therefore to treat the next slice as equivalent to a fiber facet upon which this field is impinging, and recouple the field perturbed by the complicated amplification back to the waveguide using the decomposition principal of eq. (2). This principle is illustrated in fig. 3. In consequence, the transmission matrix of each fiber slice is constructed in the following manner, element by element: the entry at some column 'c' and some row 'r', corresponding to the contribution from an incoming mode $\psi_c^s$ to an outgoing mode $\psi_r^s$, is determined as:

$$TM_{[r,c]}^{z \to z+\Delta z} = e^{-j\beta_c \Delta z}\left\{ \delta_{rc} + \tfrac{1}{2}\beta_0 \Delta z N_d (1 - jK_{RIC}) \iint \left[\sigma_e^{sig} n_2(r,\theta,z) - \sigma_a^{sig} n_1(r,\theta,z)\right] \cdot \psi_r^s(r,\theta) \cdot \psi_c^{s*}(r,\theta)\, dr d\theta \right\} \quad (6)$$

where the spatial-overlap integral arises from the decomposition of the amplified field of eq. (5) upon all supported signal modes in the fiber, and the Kronecker delta replaces the unity term in the same equation so as to reflect, for the diagonal TM elements, the passive propagation of the incoming mode $\psi_c^s$. It is worth noting that, contrary to other models treating propagation through multimode fibers in the presence of a randomization process [30-31], our model takes into account – thanks to the decomposition principle – both the guided and the non-guided components of the signal light potentially arising from the interaction with the disorder.

Finally, the total gain-dependent Transmission Matrix is readily obtained by concatenating the effects of all slices, i.e. multiplying all the 'local' TMs:

$$TM_{Total}(L, Pump) = \prod_{z'=0}^{z'=L} TM^{z'} \quad (7)$$

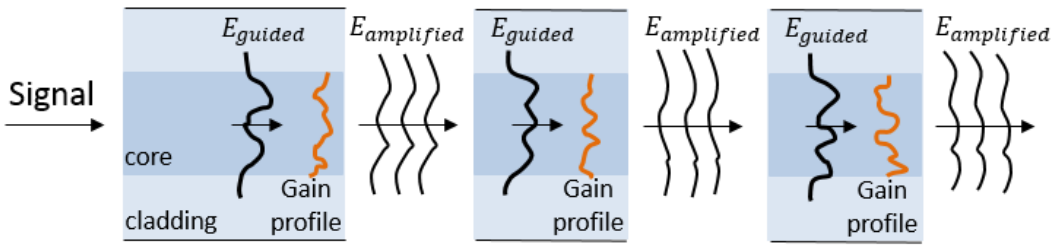

Figure 3 - Illustration of the principle of modeling the MMF amplifier as a series of slices with spatially-complex gain. The signal field that has undergone such gain must be coupled to the next slice by treating the latter as a fiber facet, i.e. by decomposing the field upon all supported fiber modes.

Our numerical simulation of the model is thus implemented: an input pumping configuration, i.e. some chosen modal content for the pump, is propagated from the injection point throughout the fiber slices in an iterative manner. In each slice, the level excitation caused by the arriving pump field is calculated as per eq. (3); the way this excitation acts back upon the pump is then taken into account by calculating the modified field, i.e. the one perturbed by the local absorption as per eq. (4), and lastly propagating said modified field to the next slice according to the principle of decomposition upon all supported pump modes. The resulting data, giving the upper level excitation for every point within the fiber volume, is then used to calculate the signal TM according to eq. (6) and (7).

*Dependence of the TM upon the Pump's Modal Content*

The theoretical model uncovers three important mechanisms which restrict the dependence of the amplifier's transmission upon the choice of pumping configuration, i.e. upon the modal content of the injected pump: the effect of upper-level excitation saturation, the scalar nature of the excitation, and the effect of the modal dispersion. The following section describes, on a qualitative level, the insights obtained regarding these limitations; additionally, results from a numerical implementation of our model for a specific case are given as quantitative demonstrations. The simulated case was an Yb-doped step-index MMFA with a core diameter of $D = 15\mu m$ and a numerical aperture $NA = 0.2$, parameters which yield guidance of about 20 modes per polarization. The pump was modeled as a coherent source of power 500mW at a wavelength of 980nm, and the signal wavelength was 1030nm. The relevant physical parameters for Ytterbium absorption, emission, level lifetime, and complex susceptibility, were taken from [25, 28-29].

I. *Limitation due to Saturation of the Pump Absorption*: The dependence upon the pump's modal content enters the picture through the spatial heterogeneity of the gain profiles; the crucial term is the variation, in the transverse coordinates, of the population-inversion within the overlap integral governing eq. (6). Pump powers strong enough to saturate this term, rendering it homogenous, will contribute to the overall gain- but not to the gain's dependence on the pump shaping. In other words, the 'control' effect we are seeking will appear mostly where highly-contrasted gain profiles emerge, i.e. in that part of the amplifier where the pump power is not exceedingly high (yet, still high enough to induce appreciable gain). This is illustrated by the excitation profiles in the insets of fig. 4, showing typical cases of the three regimes: saturation with low dependence upon modal content, emergence of spatial non-uniformity where the desired 'control' may take effect, and decay of the excitation with the absorption of practically all pump power.

An interesting quality of this limitation is that it is determined by the choice of the gain element, and more specifically - by the ratio between the emission cross-section for the signal and that of absorption for the pump: $\sigma_e^{sig}/\sigma_a^{pump}$. The qualitative reasoning behind this term is, that the degree of control could be maximized by extending the length of amplifier where non-saturated gain profiles appear - that is, by increasing the typical length of pump absorption $1/(N_d \cdot \sigma_a^{pump})$ and thus 'slowing down' the transition between saturation and complete absorption of the pump power; while at the same time requiring strong amplification of the signal over this limited length. A non-trivial point should be noted about the inability to overcome this limitation by changing the doping concentration: Although a lower concentration would indeed lengthen the segment of heterogeneous gain, it would equally reduce the amplification per unit length. This is demonstrated in the bottom panel of fig. 4, where the differential gains per unit length are plotted. The differential gain is defined as the variation in the signal transmission brought about by modification of the pump modal content; in practice it was evaluated by calculating the variation, over the ensemble of 500 randomly-drawn pumping configurations, of the elements of the 'local TM' of each slice, as per eq. (6). Comparison between the evolutions for the two concentrations demonstrates that change in $N_d$ would only serve to scale the amplifier behavior along the z axis, leaving the total degree of dependence upon our shaping ratio to be fully dictated by the cross-sections characterizing the gain element.

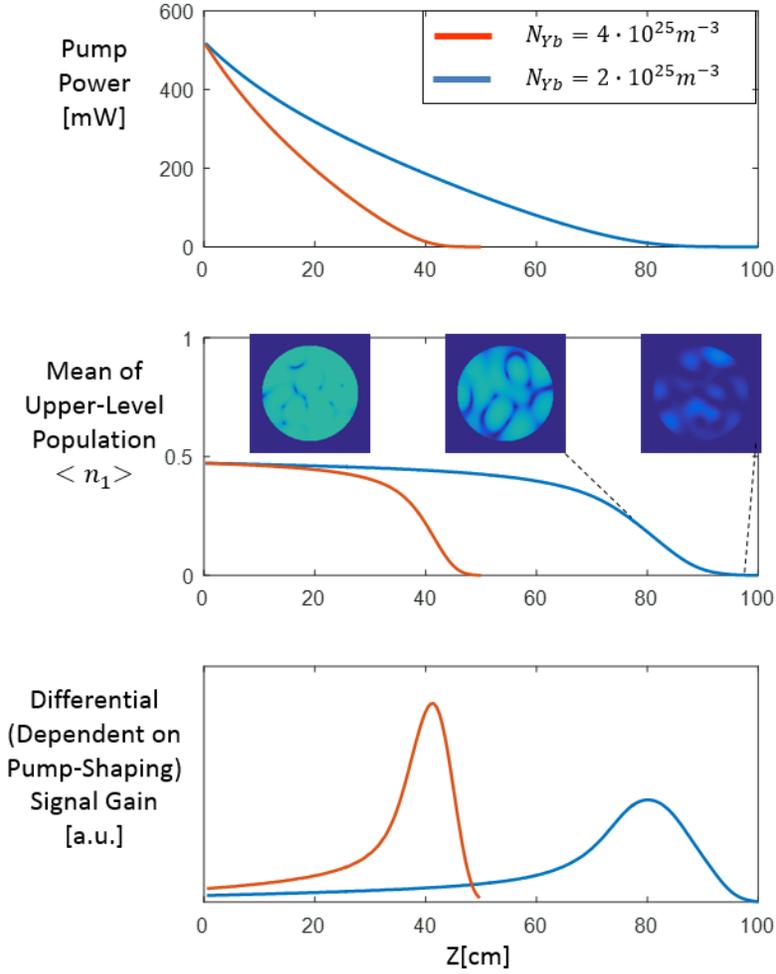

Figure 4 – Evolution of the pump power, and the resulting medium excitation along the length of a MMFA, averaged over an ensemble of 500 random pump configurations. Results are shown for two different $N_d$ values in order to demonstrate how altering the doping concentration cannot overcome the limitation upon the total accumulated 'control', and amounts only to a rescaling of the evolution in the z coordinate. Upper panel: pump power; Middle panel: upper level's excitation mean over the fiber cross-section. The insets demonstrate typical population profiles; Lower panel: the 'differential gain', as defined above.

II. *Limitation due to the Scalar Nature of the Excitation*: The interaction of the pump with the signal is an incoherent process; it is mediated through the level excitation – a physical quantity described by the population difference, a scalar number between 0 and 1, and depending only on the pump intensity. Because the gain process does not retain the pump beam's phase, no "orthogonality" exists between the degrees of freedom of our excitation (i.e. the pump modes), and those of the output we wish to control (i.e. the signal modes). Any gain profile we may choose to tailor with our pump shaping will produce some finite gain for *all* the signal modes. An example of this limited selectivity is given in fig. 5, where the transmission matrices induced in a single fiber slice of length $\Delta z = 10 \mu m$ by two different pumping configurations – each one consisting of a single pump LP mode – are compared. The pump power has been chosen so as to maximize the spatial contrast of the gain profile in the section, i.e. avoid the saturation previously discussed. The limited ability of the choice of the pump's modal content to favor a specific signal mode and suppress others is evident: although the signal mode which best matches the pumping mode is the one with the highest gain, the amplification of most other modes remains far from zero.

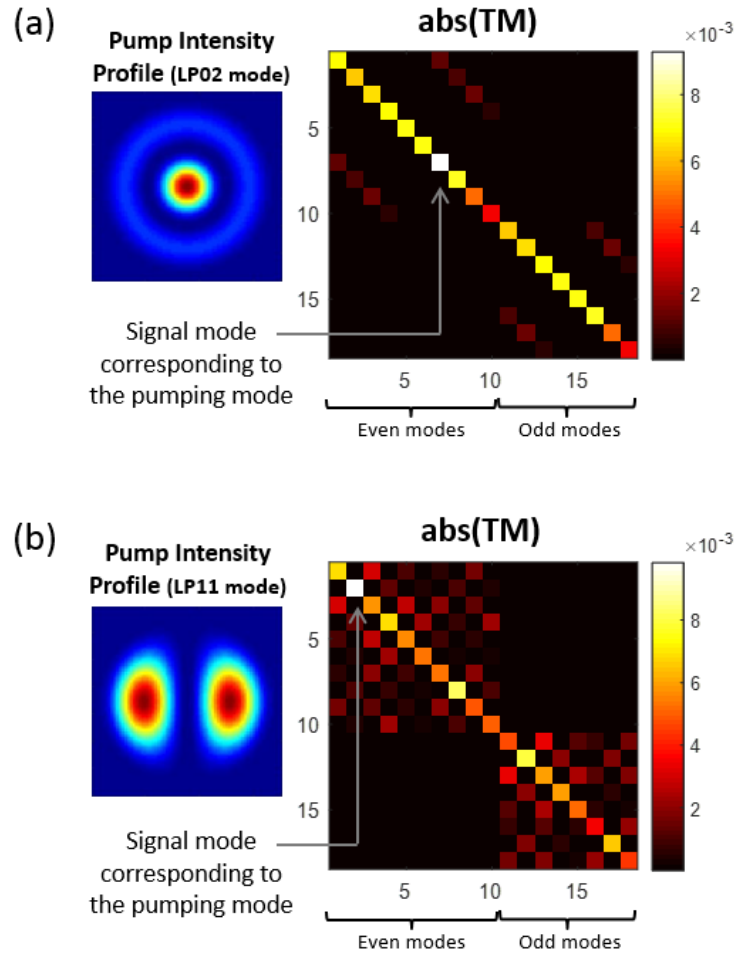

Figure 5 – The TM of a single MMFA 'slice' pumped by (a) a pure LP02 mode; (b) a pure LP11 mode. The colormap for the TM corresponds to the absolute values of its elements. The matrices are arranged to represent first the block of all 'even' modes, i.e. those whose field distribution has an angular dependence governed by a $\cos(l\theta)$ term, then the block of all 'odd' modes, i.e. those whose dependence follows a $\sin(l\theta)$ term. The TM element corresponding to the mode with the same $l, m$ indices as those of the pump mode, and hence having the best spatial overlap with the gain profile, is indicated by the arrows. These results serve to quantify the finite selectivity of the pump shaping, due to the scalar nature of the gain process.

III.     *Evolution of the Excitation Along the Fiber (Modal Dispersion)*: By wavefront shaping the pump, one may control the gain profile excited at some unique plane – for example, the injection plane at z=0. However, as the pump propagates from this plane further down the fiber, its intensity profiles will undergo the complex evolution predicted by eq. (1) (except in the unique cases where the pumping configuration is a single pure eigenmode of the fiber). As a result, the total transmission seen by the signal is the compound result of many "realizations" of pump speckles, generated at different cross-sections of the amplifier due to the modal dispersion relations. As an order-of-magnitude estimate, a typical optical fiber guiding a few tens of modes will exhibit a dephasing length (between the highest-order mode and the fundamental one) in the range of at most a few tens of microns. Considering as well the propagation lengths typically needed, in such amplifiers, in order to accumulate meaningful gain, which are at least a few tens of centimeters, it is evident that the typical number of 'slices' representing different gain profiles to be concatenated (as indicated in eq. (7)) along the fiber, is in the order of a few thousands. Evidently, the complex, disordered speckle evolution throughout the length leads to some amount of self-averaging of the spatial-overlap effects we wish to harness for the purpose of manipulating the signal's transmission.

One aspect of this averaging over length is the convergence of the amplifier TM - as the contributions are accumulated along z - towards an almost perfectly diagonal matrix; the off-diagonal elements, representing coupling between the eigenmodes, tend to stay small, albeit not completely negligible for at least some part of their population. This is demonstrated by the histograms in fig. 6a. An additional interesting insight is observed when looking at the complex values on the diagonal elements of the TM – those representing the self-amplification of each mode. By comparing the gain-induced phases to the amplitudes, a departure from a simple behavior of isolated (non-coupling) eigenmodes is revealed: perfect eigenmodes would be amplified with the dephasing/gain relation completely linear and determined by the RIC factor, as in a SM amplifier [28-29]. Whereas in our system pumping configurations may be found, for which significant deviation from the linear trendline occurs, as demonstrated by the scatter plots of fig. 6b. This disassociation between the gain-induced amplitude and phase demonstrates the importance (despite the averaging along z effect) of accounting in the MMFA model for the full TM containing the cross-contributions between all guided modes (as opposed to a simpler model, e.g. of [12-15], where the eigenmodes of the passive fiber are still considered to be the eigenmodes of the pumped amplifier, each one influenced only by its independent self-amplification).

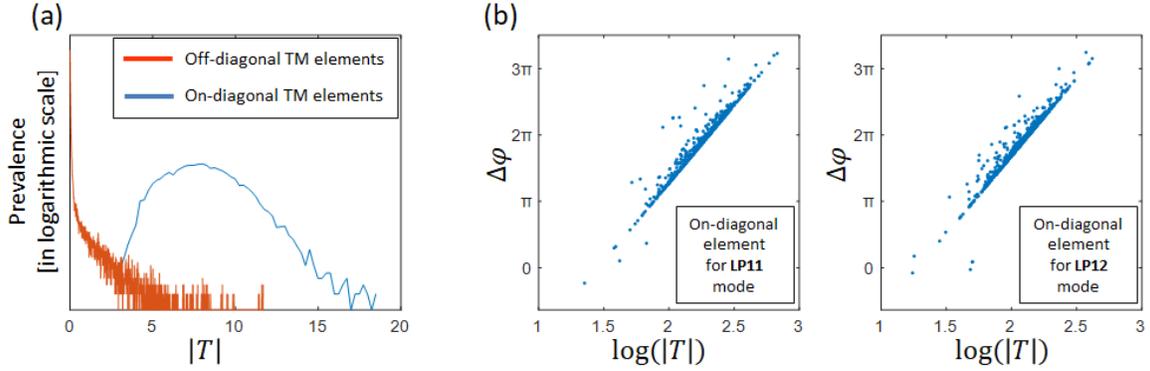

Figure 6 – Evaluation of the sensitivity of the signal TM to pump shaping, based on the statistical variation of TM elements over an ensemble of 500 random pump configurations. (a) The distribution, over the entire ensemble, of the absolute values of TM elements. In blue – the on-diagonal population, in red - off-diagonal population; (b) Dephasing vs. gain for two different signal modes, LP11 and LP12, over the ensemble. Each pump configurations is represented by a single scatter point. The departure from an ideal linear trendline is brought about only due to the effects of mode coupling.

Summarizing the section, the question naturally arising is - what degree of control of the output signal, by intelligent selection of the pump's modal content, "survives" the equalizing/averaging effects of the mechanisms discussed above? On one hand, considering the spreads along the horizontal and vertical axes of the scatter plots of fig. 6b, the sizeable range of achievable gain-induced de-phasing (spanning $2\pi$ and more) leads to hope that significant manipulation of the amplifier's output intensity pattern should be attainable. On the other hand, it is evident that full control of all the signal's degrees of freedom – meaning, simultaneously setting independent phases and amplitudes for each of the signal modes – is out of reach, because the pumping configuration interacts with all signal modes in a non-separable manner. As a preliminary assessment of the degree of tailoring of the signal output pattern achievable through pump shaping, we use once again the ensemble of transmission matrices generated by random drawing of 500 pump configurations (as previously described for fig. 4 and 6), and perform a statistical analysis on the possible variation of the output speckle pattern. We compare the TMs pair-wise, meaning that the variation is calculated as the difference between output patterns produced by two matrices, representing two different pump configurations but receiving the same input signal. A single speckle grain is chosen as a potential focus spot, and the difference in confinement of the optical power at that spot, between the TM pair, is taken to be the metric of pump-induced variation. As can be seen in the results shown in fig. 7, a change of 15% in focus confinement is rare but should be possible to find, just by random drawing of different pump modal configurations; giving reason to believe a stronger effect of focusing could be achieved with a deliberate optimization process.

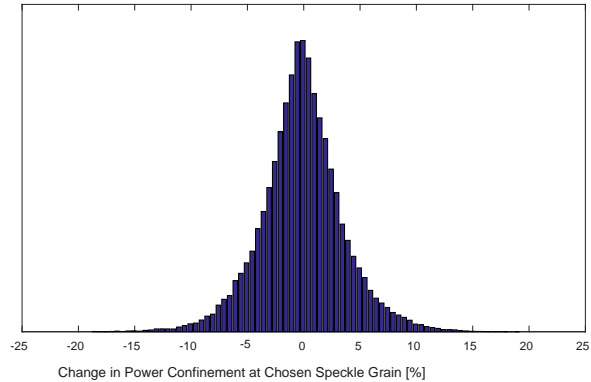

Figure 7 – Evaluation of the sensitivity of the amplifier's output speckle to pump shaping, based on the statistical variation of the power confinement at a chosen speckle grain, over an ensemble of 500 random pump configurations (measured by comparing the configurations pair-wise).

*Control of the Signal in the Spatial Domain by Pump Optimization*

Viewed as an optimization problem, control of the MMFA by selection of pumping configuration is conspicuously difficult, containing at least several tens of degrees of freedom (twice the number of guided pump modes) which interact in a highly non-linear manner to produce the signal TM. The non-linearity ensues not only from the saturation function of the medium excitation, but more importantly from the fact that the pump's intensity, rather than its field, is the excitation's driving factor. Since the gain profile of a superposition of pump modes is by no means the sum of the gains induced by the individual modes, we may not search for an optimal configuration by a separation of degrees-of-freedom approach. This is in stark contrast to optimization in a passive, linear system, as in [32], where one finds a global maximum reachable by monotonous ascent. The question of control was therefore addressed using a Global Optimization (GO) approach; for details see the materials and methods section. The GO engine received as input the MMFA parameters and the modal content of an injected signal beam, as well as the definition of a particular desired target in terms of the amplifier's transmission – for example, the focusing of the output to a single speckle grain. The results of two optimization runs, identical in the defined target (focusing at the same spot at the output), but differing by the initial random signal injected at the amplifier input, are displayed in fig. 8. As can be seen, the optimization yielded improvements of almost 23% and 20%, respectively, for the two runs. These results indeed exceed the typical confinement changes obtained in fig. 7 by random drawing of pump configurations, while evidently demonstrating that the control of the output pattern is not complete. Based on the above discussion of inherent limitation and averaging mechanisms, inability to achieve complete focusing is of course expected; however, the interesting conclusion is that considerable convergence towards a focus is clearly demonstrated.

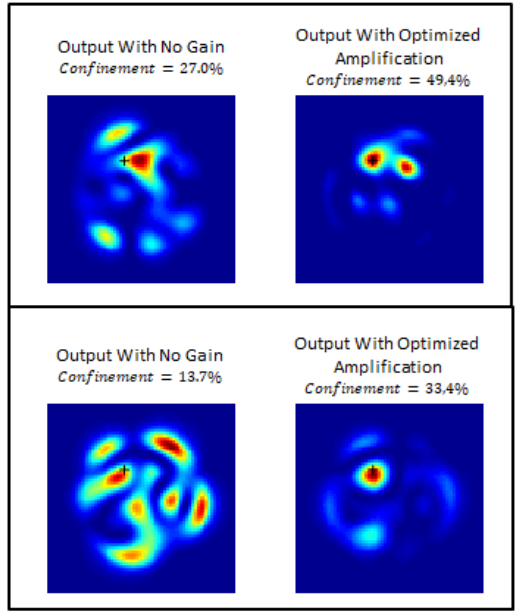

Figure 8 – Results of two different runs of the optimization algorithm, set to find the optimal pumping configuration for focusing the amplifier output intensity at the point marked by the dark + sign. On the left are the speckle patterns at the output in the absence of pumping; on the right are the patterns generated by the best (i.e. the most 'focusing') pumping configurations found by the optimization. The confinement metric is defined is defined as the percentage of optical power, out of the total guided power, found in the speckle grain chosen as the focus point.

**Discussion**

The work explored the propagation of light in a multimode amplifying fiber (MMFA), and the outlooks offered for control of its output pattern by the multimodality of the pump – more specifically, the ability to choose the modal composition of a coherent pump light injected into the system by means of wavefront shaping techniques. A theoretical model for evaluating the gain-dependent TM of the signal in such a system was proposed, taking into account the potential coupling of the stimulated emission contributed to the signal at each fiber cross-section by the spatially-complex gain profile excited there, into all supported signal modes. Key characteristics of the pump-medium and medium-signal interactions, which are expected to limit the extent of our desired control, were identified and discussed: the absorption saturation restricting the spatial contrast of the excitation, the scalar nature of the excitation restricting the selectivity in terms of the signal modes, and the self-averaging brought about by the complicated evolution of the intensity patterns along the fiber length, due to the modal dispersion. Despite these mechanisms, the interesting predictions of a numerical implementation of the MMFA model were that spatial modulation of the pump should enable significant manipulation - albeit not full control – of the signal's TM.

**Materials and Methods**

The numerical simulations of the TM of an amplifier pumped by a given pumping configuration were carried out using a Matlab implementation the MMFA model described above. The code, in the form of an easy-to-use GUI utility, is available in this link: https://github.com/tomspe/MMFAmplifier-utility .

The Global Optimization function, which searched for the optimal pumping configuration for approaching a target such as focusing the output signal, was based on a Social Spider Algorithm [33]. The algorithm works by refining a population of pump configurations, using iterative cycles of signaling between the population members in order to promote attraction towards the stronger ones. In each cycle, calculation of the different TMs corresponding to the current population was performed in parallel over a computation cluster, based on a C++ implementation of the MMFA model; then the configurations yielding the best amplified signal outputs (in terms of the target metric) were used to advance the population in the explored space.

**Acknowledgements**

This research was supported by the Agence Nationale de la Recherche N°ANR: 12-BS09-003-015. P. S. is thankful to the CNRS support under grant PICS-ALAMO. This research was also supported in part by The Israel Science Foundation (Grants No. 1871/15 and 2074/15) and the United States-Israel Binational Science Foundation NSF/BSF (Grant No. 2015694). The authors wish to thank Raphael Florentin, Vincent Kermene, Alain Barthélémy, and Agnes Desfarges-Berthelemot, of the XLIM research institute, and Jean-Pierre Huignard of the Langevin Instiue, for their support and assistance.

**Conflicts Of Interest**

The authors declare no conflicts of interest.

**References**

[1] R. Di Leonardo, S. Bianchi, Hologram Transmission Through Multi-mode Optical Fibers, Opt. Exp., Vol. 19 No. 1, 247-252 (2011).

[2] D. Loterie, S. A. Goorden, D. Psaltis, C. Moser, Confocal Microscopy Through a Multimode Fiber Using Optical Correlation, Opt. Lett., Vol. 40 No. 24, 5754-5757 (2015).

[3] S. Gigan, Viewpoint: Endoscopy Slims Down, Physics 5, No. 127 (2012).

[4] E. Desurvire, J. R. Simpson, P. C. Becker, High-gain erbium-doped traveling-wave fiber amplifier, Opt. Lett., Vol. 12 No. 1, 888-890 (1987).

[5] D. J. Richardson, J. M. Fini, L. E. Nelson, Space-Division Multiplexing in Optical Fibres, Nat. Phot., No. 7, 354-362 (2013).

[6] H. Ono, Optical Amplification Technologies for Space Division Multiplexing, NTT Tech. Rev., Vol. 15 No. 6 (2017).

[7] D. Taverner, D. J. Richardson, L. Dong, and J. E. Caplen, 158-mJ pulses from a single-transverse-mode Large-Mode-Area Erbium-Doped Fiber Amplifier, Opt. Lett., Vol. 22 No. 6, 378-380 (1997).

[8] G. Canat, Y. Jaouen, J-C Mollier, Performance and Limitations of High Brightness Er3+/Yb3+ Fiber Sources, C. R. Physique 7, 177-189 (2006).

[9] J. M. Sousaa, O. G. Okhotnikov, Multimode Er-doped Fiber for Single-Transverse-Mode Amplification, Appl. Phys. Lett., Vol. 74 No. 11, 1528-1530 (1999).

[10] J. P. Koplow, D. A. V. Kliner, L. Goldberg, Single-mode operation of a coiled multimode fiber amplifier, Opt. Lett., Vol. 25 No. 7, 442-444 (2000).

[11] J. M. Fini, Bend-resistant design of conventional and microstructure fibers with very large mode area, Opt. Exp., Vol. 14 No. 1, 69-81 (2006).

[12] E. Ip, Gain Equalization for Few-Mode Fiber Amplifiers Beyond Two Propagating Mode Groups, IEEE Phot. Tech. Lett., Vol. 24, No. 21, 1933-1936 (2012).

[13] N. Bai, E. Ip, L. Guifang, Multimode fiber amplifier with tunable modal gain using a reconfigurable multimode pump, Opt. Exp., Vol. 19 No. 17, 16601-16611 (2011).

[14] Y. Jung, Q. Kang, J. K. Sahu, B. Corbett, J. O'Callagham, F. Poletti, S. Alam, D. J. Richarson, Reconfigurable Modal Gain Control of a Few-Mode EDFA Supporting Six Spatial Modes, IEEE Phot. Tech. Lett., Vol. 26 No. 11, 1100-1105 (2014).

[15] R. N. Mahalati, D. Askarov, and J. M. Kahn, Adaptive modal gain equalization techniques in multi-mode erbium-doped fiber amplifiers, J. Light. Tech., Vol. 32, No. 11, 2133–2143 (2014).

[16] R. Florentin, V. Kermene, J. Benoist, A. Desfarges-Berthelemot, D. Pagnoux, A. Barthélémy, J. P. Huignard, Shaping the Light Amplified in a Multimode Fiber, Nat. Ligh & Sci. App., No. 6 (2017).

[17] R. Florentin, V. Kermene, A. Desfarges-Berthelemot, A. Barthélémy, Fast Transmission Matrix Measurement of a Multimode Optical Fiber with Common Path Reference, IEEE Phot. J., Vol. 10 No. 5, 7104706 (2018).

[18] R. Florentin, V. Kermene, A. Desfarges-Berthelemot, A. Barthélémy, Space-time Adaptive Control of Femtosecond Pulses Amplified in a Multimode Fiber, Opt. Exp., Vol. 26 No. 8, 10682-10690 (2018).

[19] S. M. Popoff, G. Lerosey, R. Carminati, M. Fink, A. C. Boccara, S. Gigan, Measuring the Transmission Matrix in Optics: An Approach to the Study and Control of Light Propagation in Disordered Media, Phys. Rev. Lett., No. 104, 100601-100604 (2010).

[20] J. Carpenter, T. D. Wilkinson, Characterization of Multimode Fiber by Selective Mode Excitation, J. of Light. Tech., Vol. 30 No. 10, 1386-1392 (2012).

[21] M. Plöschner, T. Tyc, T. Cizmar, Seeing Through Chaos in Multimode Fibres, Nat. Phot., 9, 529-535 (2015).

[22] D. Gloge, Weakly Guiding Fibers, Applied Optics, Vol. 10 No. 10 pg. 2252, 1971.

[23] J. A. Buck, Fundamentals of Optical Fibers, 2$^{nd}$ Edition, Hoboken, New Jersey: John Wiley & Sons, 2004.

[24] C. Barnard *et al*, Analytical Model for Rare-Earth-Doped Fiber Amplifiers and Lasers, IEEE Journal of Quant. Elec., Vol. 30 No. 8 pg. 1817 (1994).

[25] R. Paschotta *et al*, Ytterbium-Doped Fiber Amplifiers, IEEE Journal of Quant. Elec., Vol. 33 No. 7 pg. 1049 (1997).

[26] S. Foster, Complex Susceptibility of Saturated Erbium-Doped Fiber Lasers and Amplifiers, IEEE Photonics Tech. Letters, Vol. 19 No. 12 pg. 895 (2007).

[27] A. L. Barnes *et al*, Absorption and Emission Cross Section of Er3+ Doped Silica Fibers, IEEE Journal of Quant. Elec., Vol. 27 No. 4 pg. 1004 (1991).

[28] J. W. Arkwright *et al*, Experimental and Theoretical Analysis of the Resonant Nonlinearity in Ytterbium-Doped Fiber, Journal of Light. Tech., Vol. 16 No. 5 pg. 798 (1998).

[29] H. S. Chiang, J. Nilsson *et al*, Experimental Measurements of the Origin of Self-Phasing in Passively Coupled Fiber Lasers, Optics Letters, Vol. 40 No. 6 pg. 962 (2015).

[30] Y. Li, D. Cohen, T. Kottos, Coherent Wave Propagation in Multimode Systems with Correlated Noise, Phys. Rev. Lett., No. 122 pg. 153903 (2019).

[31] P. Aschieri, J. Garnier, C. Michel, V. Doya, A. Picozzi, Condensation and Thermalization of Classical Optical Waves in a Waveguide, Phys. Rev. A, No. 83 pg. 33838 (2011).

[32] I. M. Vellekoop, A. P. Mosk, Focusing Coherent Light through Opaque Strongly Scattering Media, Opt. Lett., Vol. 32 No. 16 pg. 2309-2312 (2007).

[33] J. J. Q. Yu, V. O. K. Li, A Social Spider Algorithm for Global Optimization, App. Soft Computing, Vol. 30 pg. 614-627 (2015).